\documentclass[12pt,preprint2]{emulateapj}
\usepackage{pslatex}
\usepackage[T1]{fontenc}
\usepackage[latin1]{inputenc}
\usepackage{subfigure}
\usepackage{amsmath}
\usepackage{graphicx,graphics}
\usepackage{amssymb}
\usepackage[english]{babel}

\newcommand{\gap}{\;\rlap{\lower 2.5pt \hbox{$\sim$}}\raise 1.5pt\hbox{$>$}\;}
\newcommand{\lap}{\;\rlap{\lower 2.5pt \hbox{$\sim$}}\raise 1.5pt\hbox{$<$}\;}
\newcommand{\beq}{\begin{equation}}
\newcommand{\eeq}{\end{equation}}
\newcommand{\msun}{M_\odot}
\newcommand{\mh}{M_{\rm bh}}

\newcommand{\tr}{t_{\rm r}}
\newcommand{\trh}{t_{\rm rh}}
\newcommand{\rbh}{r_{\rm bh}}
\newcommand{\tcr}{t_{\rm cr}}
\newcommand{\mnuc}{M_{\rm nuc}}
\newcommand{\mgal}{M_{\rm gal}}

\newcommand{\tgal}{t_{\rm gal}}
\newcommand{\tnuc}{t_{\rm nuc}}
\newcommand{\rgal}{r_{\rm gal}}
\newcommand{\rnuc}{r_{\rm nuc}}
\newcommand{\vgal}{V_{\rm gal}}
\newcommand{\vnuc}{V_{\rm nuc}}

\shorttitle{Nuclear Star Clusters}
\shortauthors{David Merritt}

\begin{document}

\title{Evolution of Nuclear Star Clusters}

\author{David Merritt}
\affil{Department of Physics, 85 Lomb Memorial Drive, Rochester Institute of Technology, Rochester, NY 14623\\ {\it and} \\
Center for Computational Relativity and Gravitation, School of Mathematical Sciences, 78 Lomb Memorial Drive, Rochester Institute of Technology, Rochester, NY 14623}

\begin{abstract}
Two-body relaxation times of nuclear star clusters
are short enough that gravitational encounters
should substantially affect their structure
in 10 Gyr or less.
In nuclear star clusters without massive black holes, 
dynamical evolution is a competition between core collapse, 
which causes densities to increase,
and heat input from the surrounding galaxy, 
which causes densities to decrease.
The maximum extent of a nucleus that can resist expansion is derived
numerically for a wide range of initial conditions;
observed nuclei are shown to be compact enough 
to resist expansion, although there may have been an earlier
generation of low-density nuclei that were dissolved.
An evolutionary model for NGC 205 is presented which suggests
that the nucleus of this galaxy has already undergone core collapse.
Adding a massive black hole to a nucleus inhibits core collapse,
and nuclear star clusters with black holes always expand,
due primarily to heat input from the galaxy and secondarily
to heating from stellar disruptions.
The expansion rate is smaller for larger black holes due to the 
smaller temperature difference between galaxy and nucleus
when the black hole is large.
The rate of stellar tidal disruptions and its variation with time
are computed for a variety of initial models. 
The disruption rate generally decreases with time due to the 
evolving nuclear density, particularly in the faintest galaxies,
assuming that scaling relations
derived for luminous galaxies can be extended to low luminosities.
\end{abstract}


\section{Introduction}

Many galaxies have compact stellar nuclei.
A nearby example is NGC 205, the dwarf elliptical (dE) 
companion to the Andromeda galaxy.
NGC 205 exhibits a sharp upturn in its
surface brightness profile inside of a few arcseconds, 
attributable at least in part to a population of young stars 
near the center \citep{Hodge:73,CS:90}.
The mass of the nucleus is $\sim 10^6\msun$
and its radius is $\sim 10$pc
\citep{Jones:96,Valluri:05}.
\cite{Reaves:77} and \cite{RSS:77} found similar central 
light excesses in many early-type galaxies in the Virgo cluster; 
\cite{BTS:87}, in their comprehensive survey of Virgo, found that
roughly one-fourth of the dwarf (dE and dS0 type) galaxies were nucleated.
The nuclei in the Virgo galaxies are sufficiently compact that
they appear unresolved in ground-based images, and
their detection in these photographic surveys
was facilitated by the low central surface brightnesses
of dwarf galaxies.

Recent observations with the Hubble Space Telescope have revealed
that nuclear star clusters are present in a much wider variety
vof galaxies, including luminous elliptical galaxies and 
the bulges of spiral galaxies \citep{carollo97,carollo98,boker02,ACS8}.
The fraction of galaxies with clearly distinguishable nuclear
clusters is now believed to be as high as 50\%-75\%.
The nuclei are not present in E galaxies brighter than 
absolute blue magnitude $M_B\approx -19$, 
but this may be due to the steeply-rising luminosity profiles
in these galaxies which would make detection of a nuclear excess difficult.
The frequency of nucleation also appears to drop in the
faintest galaxies, falling essentially to zero at galaxy absolute
magnitudes $M_B\approx -12$ and below \citep{SBT:85,VDB:86}.

Nuclear star clusters have the following properties
\citep[][and references therein]{Boeker:07,VDM:07}.
(a) They are 10-100 times brighter than typical Milky Way
globular clusters.
(b) Their sizes correlate with their luminosities as
$r\sim L^{0.5}$.
(c) Spectra reveal extended star formation histories, and
the luminosity-weighted mean stellar age correlates with 
galaxy Hubble type, from
$\sim 10$ Myr in late-type spirals to $\sim 10$ Gyr in ellipticals.
However the mass is typically dominated by an old stellar
population.
(d) Nuclear masses correlate with host galaxy (bulge)
masses as $M_{\rm nuc}\approx 0.003 M_{\rm bulge}$, as
well as with other global properties of the bulge
such as velocity dispersion and S\'ersic index.

The fact that nuclear star clusters obey similar scaling 
relations with host galaxy properties as do supermassive
black holes has led to the suggestion that there may
be a fundamental connection between the two classes of object
\citep{ferrarese06a,wehner06}: perhaps both should be
assigned to the single category of ``central massive object''.
This suggestion raises the question of whether nuclear
star clusters can co-exist with supermassive black holes.
\cite{Seth:08} list roughly a dozen galaxies with nuclear
emission lines or X-ray luminosities indicative of an active nucleus,
and which also contain nuclear star clusters.

This paper addresses the evolution of nuclear star clusters due to
random gravitational encounters between stars.
Ignoring the possible presence of a massive black hole,
relaxation times in the nuclei work out to be about an order 
of magnitude longer than those in globular clusters, 
but still short enough that
gravitational encounters would substantially affect
their structure after 10 Gyr.
For instance, in NGC 205, which has one of the best-resolved
nuclear star clusters, the central relaxation time is only $\sim 10^8$ yr, 
short enough that core collapse could have already occurred
\citep{Jones:96,Valluri:05}.

As pointed out by \cite{DO:85}, \cite{Kandrup:90}
and \cite{Quinlan:96}, evolution of a star cluster at the
center of a galaxy can be qualitatively different than that 
of an isolated stellar system.
If the nucleus is sufficiently diffuse, its velocity dispersion
will be lower than that of the surrounding galaxy, and
the usual tendency toward core collapse will be opposed by
a flow of energy from galaxy to nucleus.
The fate of the nucleus -- collapse versus expansion -- depends
on whether the core collapse time is short or long compared
to the time for the nucleus to absorb energy from the galaxy.
\cite{Quinlan:96} analyzed a simple family of two-component
models and derived a condition for
the minimum compactness of a nucleus in order to resist expansion.
Quinlan's condition implies a maximum size 
$\rnuc/\rgal \approx 0.02$ for nuclei with masses 
$\mnuc/\mgal=10^{-3}$.
This is intriguingly close to the observed sizes of nuclear
star clusters.

Even a nucleus compact enough to satisfy Quinlan's criterion
could escape core collapse if it contained a black hole.
A black hole also acts as a heat source, by destroying
and absorbing stars \citep{Shapiro:77}.
Nuclear star clusters containing massive black holes
should therefore expand, regardless of their initial
degree of compactness.

In this paper, integrations of the orbit-averaged Fokker-Planck
equation are carried out for models of nucleated galaxies
with and without black holes.
The Fokker-Planck model includes a loss term describing scattering
of stars into the tidal disruption sphere of a black hole,
if present.
Physical collisions between stars are ignored, and the
mass liberated from disrupted stars is assumed to escape 
instantaneously from the galaxy; these are both approximations
but not unreasonable ones (e.g. in NGC 205 the timescale
for physical collisions is $\sim 3$ orders of magnitude longer
than the relaxation time).
Initial conditions are derived from mass models that mimic
the observed luminosity profiles of nucleated galaxies.
Quinlan's criterion for black-hole-free nuclei is refined;
we find that observed nuclei are either in the core-collapse
regime, or have sufficiently long relaxation times that they
would not have evolved appreciably in 10 Gyr.
In models containing black holes, the dominant 
process driving nuclear evolution is almost always found 
to be heating by the galaxy.

The models presented here are idealizations in one important
respect: only a single stellar mass group is considered and
stellar masses are assumed to be fixed in time.
The effects of mass segregation, star formation and
stellar evolution are therefore ignored.
Multi-mass models will be considered in a later paper.
While the calculations presented here have some relevance to nuclear
star clusters in all galaxies, the focus is on models
that are structurally similar to low-luminosity elliptical galaxies.

Section II summarizes the properties of nuclear star clusters
and their host galaxies that are relevant to this study.
Sections III and IV describe the results of evolutionary calculations
without, and with, black holes, respectively.
Section V sums up.

\begin{figure*}
\centering
\includegraphics[scale=0.80,angle=-90.]{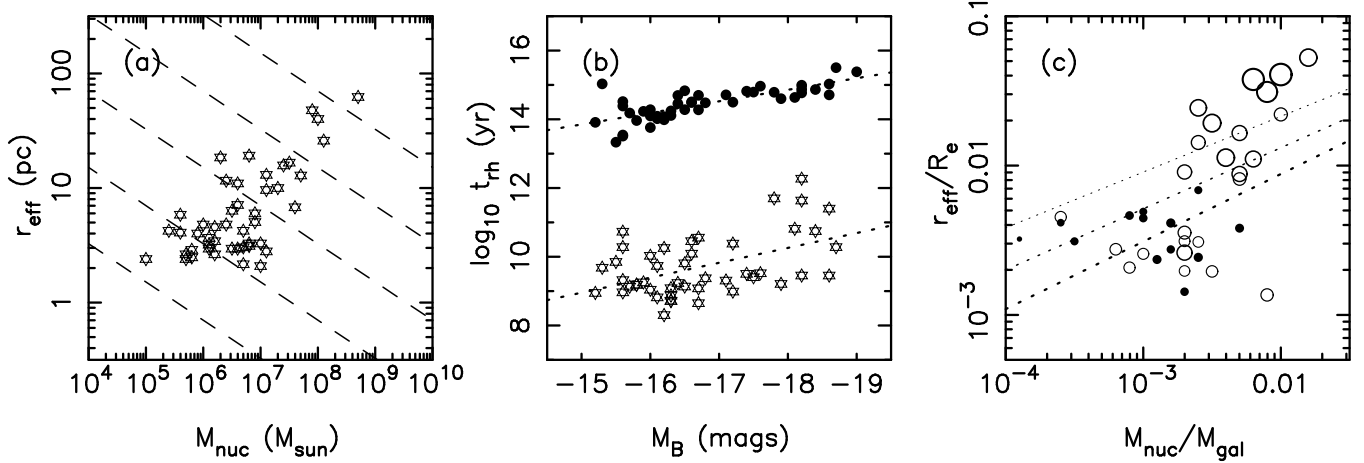}
\caption{Properties of nuclear star clusters in the subset of 
Virgo cluster early-type galaxies that were found to be
nucleated in the ACS Virgo Cluster Survey \citep{ACS1}.
(a) Nuclear radii and masses; masses are from the compilation
of Seth et al. (2008). Dashed lines correspond to nuclear half-mass
relaxation times of $(10^8,10^9,10^{10}, 10^{11},10^{12})$ years
increasing up and to the right.
(b) Half-mass relaxation times of nuclei (stars) and their
host galaxies (filled circles).
Dotted lines are least-squares fits.
(c) Vertical axis is the ratio of half-light radii of nuclei
($r_{\rm eff}$) to half-light radii of galaxies ($R_e$);
horizontal axis is the ratio of nuclear mass to galaxy mass.
Symbol size is proportional to the logarithm
of the nuclear relaxation time in (b).
Open circles have $\log_{10} (t_{rh}/{\rm yr})\ge 9.5$
and filled circles have $\log_{10} (t_{rh}/{\rm yr})<9.5$.
Dashed lines indicate the critical value of $r_{\rm eff}/R_e$
above which nuclei expand, for galaxies described by
Einasto indices $n=2$ (top), $n=3$ and $n=4$ (bottom).
}
\label{fig:nuclei}
\end{figure*}

\section{Nuclear Star Clusters and their Host Galaxies}

\subsection{Nucleated Galaxies in the Virgo Cluster}

Figure~\ref{fig:nuclei}a shows half-light radii $r_{\rm eff}$
and masses $\mnuc$ of nuclei observed in the ACS Virgo
Cluster Survey \citep{ACS1}.
Dashed lines indicate constant values of the half-mass 
relaxation time
\beq
t_{rh} = {1.7\times 10^5 \left[r_h({\rm pc})\right]^{3/2}N^{1/2}\over 
\left[m/M_\odot\right]^{1/2}} \ {\rm years}
\label{eq:tr}
\eeq
\citep{Spitzer:87},
with $r_h$ set equal to 
$r_{\rm eff}$.
The mean stellar mass $m$ was set to one solar mass and 
$N$ was computed from $N=\mnuc/m$. 
Half-mass relaxation times for many of the nuclei are less
than 10 Gyr.

Figure~\ref{fig:nuclei}b shows the dependence of relaxation time
on galaxy absolute blue magnitude $M_B$.
Relaxation times are plotted both for the nuclei, and for their
host galaxies; in the case of the galaxies, $r_h$ in equation~(\ref{eq:tr})
was equated with the observed effective radius $R_e$ 
(e.g. Ciotti 1991) and $N$ was set to $\mgal/m$.
Dotted lines are least-squares fits to $\log t_{rh}$ vs. $M_B$
for galaxies and nuclei independently; the fitted relations are
\begin{mathletters}
\begin{eqnarray}
\log_{10} (\tgal/\rm{yr}) &=& 14.2 - 0.336 (M_B+16), 
\label{eq:leastsquares}
\\
\log_{10} (\tnuc/\rm{yr}) &=& 9.38 - 0.434 (M_B+16)
\label{eq:leastsquares2}
\end{eqnarray}
\end{mathletters}
where we have written $\tgal$ and $\tnuc$ for the half-mass
relaxation time of the galaxy and the nucleus respectively.
The second of these relations predicts nuclear relaxation times
below $0.1$ Gyr in galaxies with $M_B\approx -12$, the least
luminous galaxies that contain compact nuclei \citep{VDB:86}.

Figure~\ref{fig:nuclei}c plots nuclei in the size ratio
vs mass ratio plane.
The symbol sizes in this plot are proportional
to $\log_{10}\tnuc$; filled symbols have
 $\tnuc<3$ Gyr and open symbols have  $\tnuc>3$ Gyr.
There is evidence in this figure for two populations:
compact nuclei with short relaxation 
times and more extended nuclei with long 
relaxation times.
Figure~\ref{fig:nuclei}a, combined with the knowledge that nuclear
masses correlate with galaxy masses \citep{wehner06,ferrarese06a},
implies that the large-$\tnuc$ population 
in Figure~\ref{fig:nuclei}c is associated
with the brightest galaxies.
The lack of nuclei with $r_{\rm eff}/R_e\lap 10^{-3}$
probably reflects an observational limitation:
at the distance of Virgo, the ACS/WFC pixel scale of
$\sim 0.049$ arc second corresponds to $\sim 3.9$ pc,
or $\sim 0.004$ times a typical galaxy effective radius
of $\sim 1$ kpc.

\begin{figure}
\centering
\includegraphics[scale=0.45]{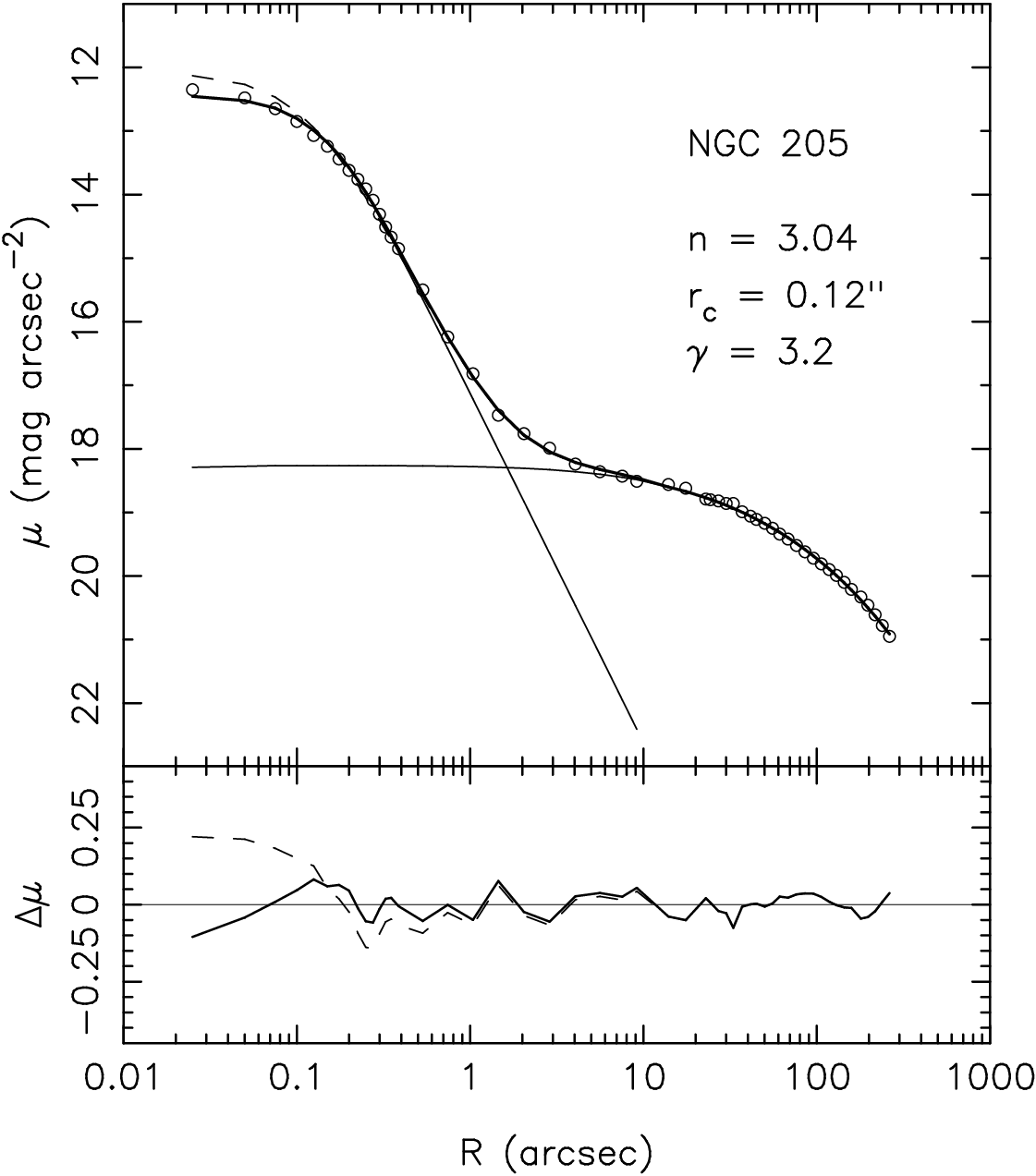}
\caption{Two-component (eq.~\ref{eq:model}) fit to surface 
brightness data of NGC205.
Thin solid lines show the nuclear and galactic components
separately, after convolution with the ACS point-spread function;
dashed line is before PSF convolution.
Lower panel show surface-brightness residuals.
$n$: Einasto index of the galaxy component;
$r_c$: nuclear core radius;
$\gamma$: power-law index of the nuclear space density.}
\label{fig:n205}
\end{figure}

\begin{figure}
\centering
\includegraphics[scale=0.60]{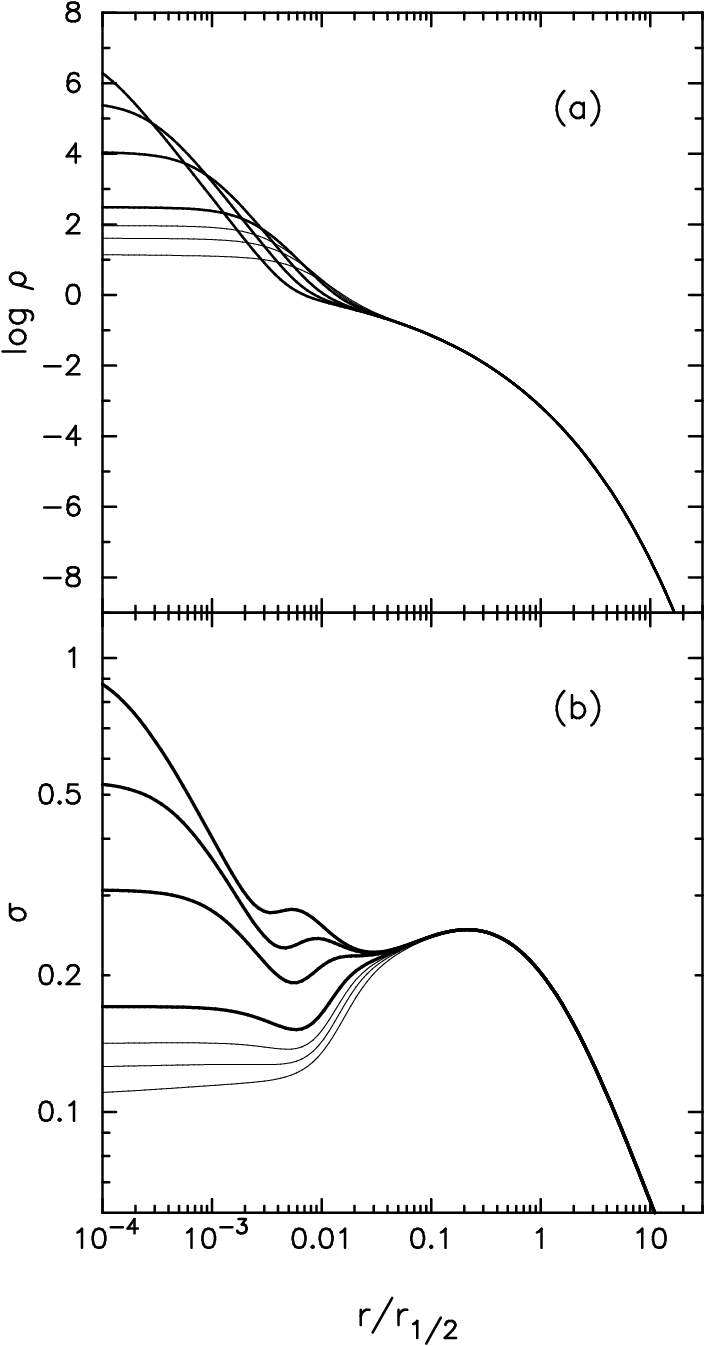}
\caption{Density (upper) and velocity dispersion (lower) profiles
of a set of galaxy models constructed according to 
equation~(\ref{eq:model}), normalized such that $G=M_{\rm gal}=1$.
All models have $M_{\rm nuc}/M_{\rm gal}=0.003$,
$n=3$ and $\gamma=4$.
Different curves correspond to different degrees of nuclear compactness,
i.e. $\log_{10}(r_{\rm nuc}/r_{\rm gal})=
(-3.72,-3.20,-2.72,-2.20,-2.02,-1.90,-1.72)$.
Thick lines denote models in which the nucleus undergoes
``prompt'' core collapse; thin lines are models in which heat transfer
from the galaxy causes the nucleus to expand initially
(as show in Fig.~\ref{fig:evol}).}
\label{fig:two}
\end{figure}

\begin{figure*}
\centering
\includegraphics[scale=0.65,angle=-90.]{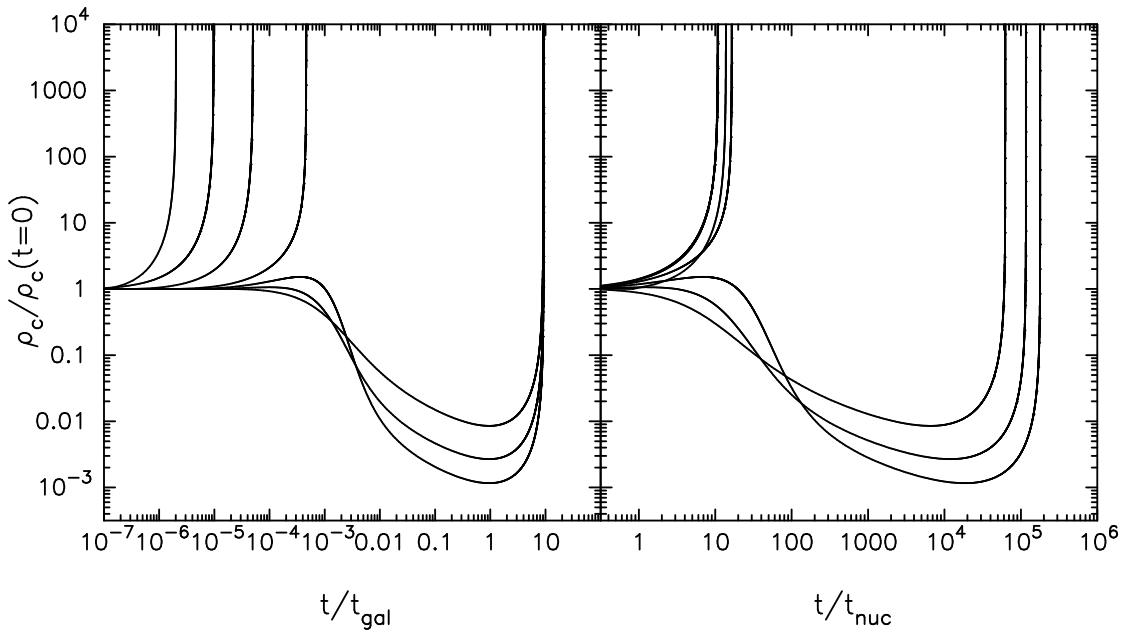}
\caption{Evolution toward core collapse in the galaxy models
plotted in Fig.~\ref{fig:two}.
Vertical axis is the core density normalized to its initial
value.
The horizontal axis is the time in units of the initial,
half-mass relaxation time $t_{rh}$, defined by the galaxy
(left) or by the nucleus (right).}
\label{fig:evol}
\end{figure*}

\subsection{NGC 205}

Nuclear star clusters beyond the Local Group are too distant
for their structure to be well resolved.
The nearby ($\sim 824$ kpc; McConnachie et al. 2005) dE galaxy
NGC 205 contains one of the best resolved nuclei.
Figure~\ref{fig:n205} shows surface brightness data
of NGC 205 in the $I$ band,
obtained by combining $HST$ ACS observations  
from \cite{Valluri:05} with 
ground-based data of \cite{Kim:98} and \cite{Lee:96}.
The horizontal axis is $r_{\rm SMA}\left[1-\epsilon(r)\right]^{1/2}$
where $r_{\rm SMA}$ is the isophotal semi-major axis and $\epsilon$
is the ellipticity; NGC 205 is moderately elongated,
$\langle\epsilon\rangle \approx 0.5$.

These circularly-symmetrized data were fit to a two-component 
spherical model defined, in terms of the (3d) luminosity density
$j(r)$, as
\begin{mathletters}
\begin{eqnarray}
j(r) &=& j_1(r) + j_2(r) \\
j_1(r) &=& j_{\rm gal}
e^{-b\left[\left(r/r_{\rm 1/2}\right)^{1/n}-1\right]} \\
j_2(r) &=& j_{\rm nuc}
\left(1+r^2/r_c^2\right)^{-\gamma/2}.
\end{eqnarray}
\label{eq:model}
\end{mathletters}
The galaxy is represented by Einasto's (1965) law,
identical in functional form to the law of S\'ersic (1963) that is
more typically applied to surface density profiles.
The quantity $b(n)$ is chosen such that $r_{\rm 1/2}$
is the (3d) radius containing $1/2$ of the total mass corresponding
to $j_1$; $n$ is an index that controls the curvature of
the profile (not to be confused with the S\'ersic index;
the latter is typically smaller by $\sim 1-2$ than the Einasto $n$
that describes the space density).
The nucleus is represented by a generalized ``Hubble'' law;
$r_c$ is the core radius and $\gamma$ controls the
steepness of the nuclear density falloff, e.g. $\gamma=5$
is Plummer's (1911) model and $\gamma=2$ is the isothermal sphere.
Equation~(\ref{eq:model}) was compared to the data
of Figure~\ref{fig:n205} after spatial projection and convolution with
the circularly-symmetrized ACS HRC point-spread function.
The optimization routine adjusted the five parameters
$(j_{\rm gal}, j_{\rm nuc}, r_{\rm 1/2}, r_c, n)$
at fixed $\gamma$; the latter parameter was varied in steps
of $0.1$.

Figure~\ref{fig:n205} shows the fit that minimizes
the surface-brightness residuals, with $\gamma=3.2$.
The fit is good ($\sim 0.05$ magnitude rms deviation),
although this decomposition of the brightness profile
into ``galaxy'' and ``nucleus'' is not certainly not unique.

Since it provides such a good description of NGC 205,
the two-component model of equation~(\ref{eq:model})
will be used as a basis for the evolutionary studies
described below.
We note that one parameter in this fit, 
the Einasto index $n$ (or rather, the corresponding Sersic
index describing the surface brightness profile; Einasto
fits have not yet been carried out for many galaxies)
is known to correlate roughly with galaxy luminosity,
in the sense that brighter galaxies have larger $n$
\citep{GG:03}.

\section{Nuclear Evolution without Black Holes}

We first consider evolution, due to gravitational 
encounters, of a galaxy containing a compact nucleus
consisting of stars with a single mass and
no nuclear black  hole.
Physical collisions between stars are ignored.
In this idealized model, evolution takes place
on a time scale determined roughly by the central (nuclear)
relaxation time, $t_{rc}$; stars are scattered 
into less-bound orbits, transferring heat to the
nuclear envelope and causing the central density to
increase.
If the nucleus were isolated, and ignoring sources
of heat like hard binary stars, core collapse would
take place in a time of $\sim 10^2 t_{rc}$ 
\citep{Spitzer:87}.
However because the nucleus is embedded in a galaxy,
flow of heat can also take place in the opposite
sense, from galaxy to nucleus, causing the latter
to expand; core collapse is then delayed until
a much longer time has elapsed \citep{Quinlan:96}.

\subsection{Evolution timescales}

We begin by deriving an approximate expression for the time scale to
transfer energy from a galaxy to a nucleus, then derive an approximate 
condition for heating to reverse core collapse.
Our approach is similar to treatments in \cite{DO:85} and
\cite{Kandrup:90}.

Consider a two-component system consisting of galaxy and nucleus.
Both components are assumed homogeneous and are characterized by a 
density ($\rho_{\rm gal},\rho_{\rm nuc}$), 
radius ($\rgal,\rnuc$), 
rms velocity ($\vgal,\vnuc$), 
and half-mass relaxation time ($\tgal,\tnuc$).

Define $\epsilon_{\rm nuc}\equiv (1/2)\rho_{\rm nuc} V_{\rm nuc}^2$ to
be the kinetic energy per unit volume of the nucleus.
Assuming Maxwellian velocity distributions
and a single stellar mass $m$,
the rate of change of $\epsilon_{\rm nuc}$ due to gravitational
encounters is
\beq
{d\epsilon_{\rm nuc}\over dt} = 4\sqrt{6\pi} G^2m \rho_{\rm nuc}
\rho_{\rm gal}\ln\Lambda {\left(V_{\rm gal}^2-V_{\rm nuc}^2\right)\over
\left(V_{\rm gal}^2 + V_{\rm nuc}^2\right)^{3/2}}
\eeq
(Spitzer 1987, eq.~2-60).
A necessary and sufficient condition for the nucleus to be
heated by the galaxy is $\vgal>\vnuc$.
When this condition is satisfied, the nuclear 
heating time is
\begin{mathletters}
\begin{eqnarray}
\tau_{\rm heat} &\equiv& \left|{1\over\epsilon_{\rm nuc}}
{d\epsilon_{\rm nuc}\over dt}\right|^{-1} \\
&=& {1\over 48}{\sqrt{6\over\pi}}{\vnuc^3\over G^2 m\rho_{\rm nuc}\ln\Lambda}
\left({\rho_{\rm nuc}\over\rho_{\rm gal}}\right)
{\left(\vgal^2/\vnuc^2+1\right)^{3/2}\over
\left(\vgal^2/\vnuc^2-1\right)}.
\end{eqnarray}
\end{mathletters}
In the limiting case $\vgal\gg \vnuc$ this becomes
\begin{mathletters}
\begin{eqnarray}
\tau_{\rm heat} &=& \left({\rho_{\rm nuc}\over\rho_{\rm gal}}\right)
\left({\vgal\over \vnuc}\right) \label{eq:theata}
{1\over 48}{\sqrt{6\over\pi}}
{\vnuc^3\over G^2 m\rho_{\rm nuc}\ln\Lambda}\\
&\approx& \left({\rho_{\rm nuc}\over\rho_{\rm gal}}\right)
\left({\vgal\over \vnuc}\right) \tnuc
\approx \left({\vgal\over \vnuc}\right)^2 \tgal
\label{eq:theatc}\\
&\approx& \left({\rho_{\rm nuc}\over\rho_{\rm gal}}\right)^{1/2}
\left({\vnuc\over\vgal}\right)^{1/2} \left(\tnuc\tgal\right)^{1/2}.
\label{eq:theatd} 
\end{eqnarray}
\label{eq:theat}
\end{mathletters}
\noindent Models that satisfy the condition $\vgal>\vnuc$ have
$\rho_{\rm nuc}\gap\rho_{\rm gal}$.
Thus, the nuclear heating time is of the same order as, 
or somewhat less than, the geometric mean of $\tnuc$ and $\tgal$.
According to Figure~1b,
this time is shorter than 10 Gyr in at least some galaxies.

Heating from the galaxy will reverse core collapse 
if $\tau_{\rm heat}$ is shorter than the nuclear core collapse time.
One definition of the latter time is
\beq
\tau_{\rm cc} \equiv
\left|{1\over\rho_{\rm nuc}} {d\rho_{\rm nuc}\over dt}\right|^{-1}
= \xi^{-1}\tnuc
\eeq
where $\xi^{-1}$ varies from $\sim 10$ in the early stages
of core collapse to an asymptotic value of $\sim 300$
(e.g. Cohn 1980).
The condition $\tau_{\rm heat}<\tau_{\rm cc}$ becomes
\beq
{\rho_{\rm nuc}\over\rho_{\rm gal}} {\vgal\over\vnuc} < \xi^{-1}
\label{eq:cond1}
\eeq
where $\vgal\gg\vnuc$ has again been assumed.
This expression is similar to equation~(14) of \cite{DO:85}.
We can convert equation~(\ref{eq:cond1}) into a relation
between the quantities plotted in Figure~1 by writing
$\rho_{\rm nuc}\approx\mnuc/\rnuc^3$ and by applying the
virial theorem separately to both components, i.e.
\beq
\vnuc^2\sim {G\mnuc\over\rnuc},\ \ \ 
\vgal^2\sim {G\mgal\over\rgal};
\eeq
the former expression will only be approximately true for low-density
nuclei.
With these substitutions, the condition (\ref{eq:cond1}) becomes
\beq
{\mnuc\over\mgal} \lap 10^4 \left({\xi^{-1}\over 100}\right)^2 
\left({\rnuc\over\rgal}\right)^5.
\label{eq:cond2}
\eeq
Setting $\mnuc/\mgal=0.003$, equation~(\ref{eq:cond2}) implies
that $\rnuc/\rgal$ must be smaller than $\sim 0.05$ in order
for core collapse to occur, a value that lies within the range
of observed nuclear sizes (Fig.~1c).
While the numerical coefficients in equation~(\ref{eq:cond2}) 
should not be considered accurate, 
the general form of the relation will be verified below via
more realistic evolutionary models.

\subsection{Fokker-Planck model}

Evolution 
was modelled using the isotropic, orbit-averaged Fokker-Planck 
equation  \citep{Cohn:80}, in the implementation
described by \cite{MMS:07}.
The initial galaxy was given a density law as
defined by equation~(\ref{eq:model}); the self-consistent
gravitational potential and isotropic phase-space distribution
functions were derived from Poisson's and Eddington's formulae
respectively.
Henceforth we characterize these initial models by the two dimensionless
parameters
\beq
r_{\rm nuc}/r_{\rm gal}, \ \ \ M_{\rm nuc}/M_{\rm gal}
\eeq
where $r_{\rm nuc}$ and $r_{\rm gal}$ are the spatial
(not projected) half-mass radii of nucleus and galaxy 
respectively, and $M_{\rm nuc}$ and $M_{\rm gal}$ are the 
total masses corresponding to the two components.
Given the varied and non-unique ways in which
half-light radii are defined observationally,
we will henceforth simply identify
$r_{\rm nuc}/r_{\rm gal}$ with $r_{\rm eff}/R_e$ when comparing
models to data.
(This identification should probably only be trusted to within a
factor of two or so.)

Figure~\ref{fig:two} shows $\rho(r)$ and $\sigma(r)$ for
a set of galaxy models with Einasto index $n=3$ and nuclear
slope $\gamma=4$; $\sigma(r)$ is the 1d velocity dispersion.
Each of these models has $M_{\rm nuc}/M_{\rm gal}=0.003$ but
they differ in their degree of nuclear concentration, from
$r_{\rm nuc}/r_{\rm gal} \approx 0.0002$ to 
$\sim 0.03$.
The dependence of $\sigma$ on $r$ is complex, 
sometimes exhibiting multiple maxima;
but as the nucleus is made less compact,
the central velocity dispersion drops, and when
$r_{\rm nuc}/r_{\rm gal}\gap 0.003$
the central $\sigma$ is lower than the peak value in the 
outer galaxy.
``Temperature inversions'' like these imply a flow of heat
from galaxy to nucleus, which tends to counteract the outward
flow that would otherwise drive the nucleus toward core collapse
\citep{DO:85}.

Figure~\ref{fig:evol} shows the change with time of the central
densities of the models in Figure~\ref{fig:two}.
Times are normalized by the initial, half-mass relaxation
time; the left panel normalizes to the galactic
relaxation time $\tgal$, and the right panel to $\tnuc$.
There is a critical value of $r_{\rm nuc}/r_{\rm gal}$ above which
evolution toward core collapse is halted and the nucleus
expands.
Core collapse still occurs in these models but on
a much longer time scale, roughly ten times the galaxy half-mass
relaxation time (and far longer than galaxy lifetimes;
see Fig.~\ref{fig:nuclei}b).
The delay is due, as first shown by \cite{Quinlan:96},
to the galaxy's need to first reverse the temperature gradient
near the center by creating a large, flat core.
In the models with nuclei that contract,
core collapse occurs in $15- 20$ times the 
initial nuclear half-mass relaxation time.

Experiments like the ones illustrated in Fig.~\ref{fig:evol} 
were carried out for initial models with a range of 
$r_{\rm nuc}/r_{\rm gal}$ and $M_{\rm nuc}/M_{\rm gal}$ values,
and with Einasto indices $n=(3,4,5)$.
The critical size ratio that separates collapsing-core and expanding-core
models is plotted in Figure 1c for the three different values of $n$.
The critical values are well fit by relations of the form
\beq
{M_{\rm nuc}\over M_{\rm gal}} = 
A \left({r_{\rm nuc}\over r_{\rm gal}}\right)^B
\eeq
where
\begin{eqnarray}
n=2: && A=300,\ \ B=2.70 \nonumber \\
n=3: && A=390,\ \ B=2.45 \nonumber \\
n=4: && A=355,\ \ B=2.20 . 
\label{eq:criteria}
\end{eqnarray}
Quinlan (1996) derived a similar
criterion for double-Plummer-law galaxy models;
his relation is most similar to the one found here for
$n=2$, which is reasonable given the low central 
concentration of Plummer models.
While these results are all based on nuclei with
$\rho\sim r^{-4}$ envelopes, we note that half-mass
radii are nearly invariant to the structural changes
that occur prior to core collapse.
Hence the criteria of equation~(\ref{eq:criteria})
should apply approximately to a much broader class of models,
including models in which the nucleus evolved toward
core collapse starting from the initial conditions adopted here.

Interestingly, Figure~\ref{fig:nuclei}c shows that 
observed nuclei almost all lie in the ``prompt''
core collapse regime; the only clear exceptions are
nuclei with such long relaxation times
($\gap 10$ Gyr) that very little evolution
would have occurred since their formation.

It would be nice to go one step further and argue that
the paucity of nuclei above the critical lines in 
Figure~\ref{fig:nuclei}c is due to dynamical evolution,
i.e. that there once existed a population of low-density
nuclei that underwent expansion and that now have such
low densities that they are no longer observable
as distinct components.
This argument is valid only if dissolution time scales
for the nuclei in this region of the plot are 
shorter than $\sim 10$ Gyr.
This is true for some, but not all, possible initial conditions.
In models near but above the critical line, 
the time for the central density 
to drop by a factor $\sim 2$ is 
found to be $\sim 10^{-4}t_{\rm gal}$ for
$M_{\rm nuc}/M_{\rm gal}\lap 3\times 10^{-4}$
but increases to $\sim 10^{-3}t_{\rm gal}$ for
the more typical value
$M_{\rm nuc}/M_{\rm gal}\lap 3\times 10^{-4}$
(e.g. Fig.~\ref{fig:evol}).
The latter time is longer than 10 Gyr in almost
all galaxies.

\begin{figure}
\centering
\includegraphics[scale=0.45]{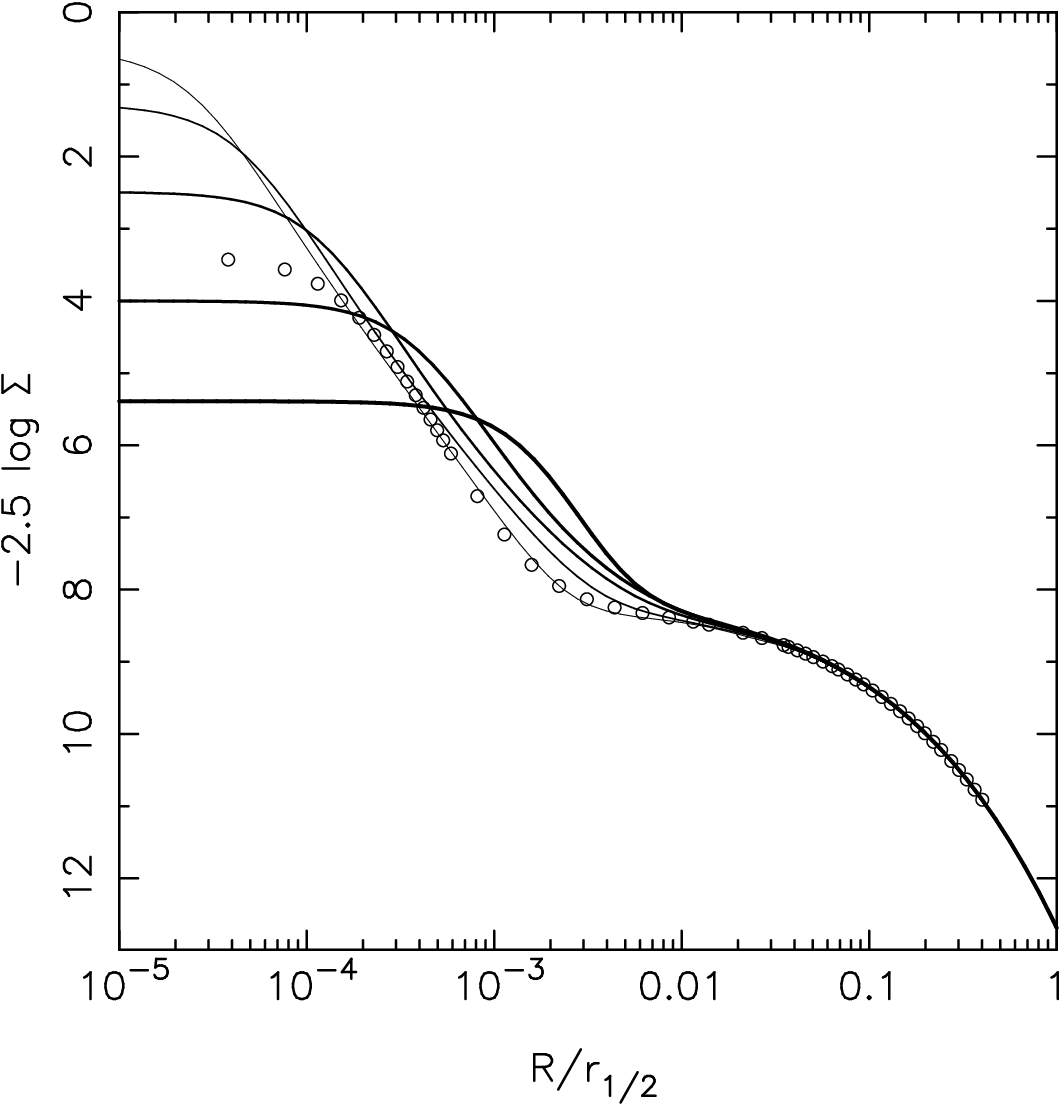}
\caption{Open circles show the surface ``mass density'' of a model
of NGC 205 derived from Fig.~\ref{fig:n205}; 
the normalization of the nuclear component
was reduced by a factor of $0.3$ relative to that of the galaxy 
to account for the lower mass-to-light ratio of the nucleus.
Heavy solid line shows a mass model that was integrated forward
via the Fokker-Planck equation;
progressively thinner lines
show the surface density profile at 
($0.8,0.98,0.997,0.9995$) times the core collapse time
of $\sim 10$ Gyr.
\label{fig:n205fp}}
\end{figure}

Finally, we consider the special case of NGC 205
(Fig.~\ref{fig:n205}).
On which side of the divide does it lie?
Dynamical modelling of NGC 205 suggests that the nuclear
mass-to-light ratio is $\sim$ a few times smaller
than that of the galaxy as a whole \citep{CS:90,Valluri:05}.
To construct a mass model, we therefore simply decreased the amplitude
of the nuclear component in the luminosity profile
fit of Figure~\ref{fig:n205}
by a factor of $0.3$ at all radii.
Figure~\ref{fig:n205fp} shows the
projected ``surface brightness'' of this mass model,
i.e., $-2.5\log \Sigma$, where $\Sigma(R)$ is the
mass surface density at projected radius $R$.
The $r_{\rm nuc}/r_{\rm gal}$ and $M_{\rm nuc}/M_{\rm gal}$ values
for this mass model place NGC 205 close enough to the
critical lines in Figure~\ref{fig:nuclei}c that its fate
is difficult to judge.

Accordingly, a variety of initial models (all with $n=3.04$
and $\gamma=4$) were constructed and integrated
forward for $\sim 10$ Gyr to see if any could match the
reconstructed mass profile of NGC 205.
Figure~\ref{fig:n205fp} shows the results of one such exercise:
this initial model undergoes ``prompt'' core collapse, achieving singular
density in a time of $\sim 1.0\times 10^{-4}$ galaxy half-mass
relaxation times, i.e. $\sim 10^{10}$ yr.
The final, projected mass profile is essentially identical
to that of NGC 205 except at $R\lap 0.1''$ where the data
suggest a flat core; this difference is probably not significant
however given that the luminosity profile is not well constrained 
at radii smaller than the PSF.
(Alternatively, NGC 205 might not have reached full core collapse;
or some process like binary heating might have caused the core to re-expand.)
While not conclusive, this and similar experiments suggest that
the luminosity profile of NGC 205 is consistent with this galaxy's
nucleus having undergone core collapse at some time in the last $10$ Gyr.

It is interesting to note that even slight reductions in the central density
of the initial model in Figure~\ref{fig:n205fp}
(larger $r_{\rm nuc}$, smaller $M_{\rm nuc}$)
led to expansion rather than collapse.
In other words: if its formation history had been only
slightly different, NGC 205 might no longer contain a compact 
nucleus.

\section{Nuclear Evolution with Black Holes}

\subsection{Models and Assumptions}

Next we consider the evolution of nucleated galaxies containing
massive black holes; the latter are idealized as Newtonian 
point masses with fixed mass $\mh$.
Evolution of $f(E)$ describing the stars
is again followed via the orbit-averaged
Fokker-Planck equation with evolving potential, 
including the (fixed) potential from the black hole,
and with a modification
that accounts for the loss of stars that are scattered
into the black hole's sphere of tidal disruption
\citep{MMS:07} 
The latter is assigned a radius of $r_t$ where
\beq
r_t=\left(\eta ^2 \frac{\mh}{m}\right)^{1/3}r_{\star}
\label{eq:rt}
\eeq
with $r_\star$ a stellar radius and $\eta \approx 0.844$, 
the value appropriate to an $n=3$ polytrope.
If $E=-\mathrm{v}^2/s+\psi(r)$ is the binding energy per unit mass
of a star in the combined potential of the galaxy
and black hole, then the tidal disruption sphere 
defines a ``loss cone'' of orbits with angular momenta 
$J$ such that $J\lap J_{lc}(E)$, where
\beq
J^2_{lc}(E) = 2r_t^2 \left[\psi(r_t)-E\right]\approx 2G\mh r_t.
\eeq
For black  hole masses of interest here ($\mh< 10^8 M_\odot$),
$r_t$ is always greater than the radius of the event horizon,
i.e. stars are disrupted not swallowed.

The Fokker-Planck equation now has the form
\beq
{\partial N\over\partial t} = -{\partial F_E\over\partial E}
- {\cal F}(E,t).
\label{eq:dNdt2}
\eeq
Here, $N(E,t)$ is the distribution of stellar energies;
${\cal F}(E,t)$ is the flux of stars per unit of energy
that are scattered in the angular-momentum direction
into the tidal disruption sphere;
and $F_E(E,t)$ is the energy-directed flux.
Each of these terms is defined in detail in \cite{MMS:07}.

Three additional, dimensionless parameters are associated 
with  the black hole models:
$r_t/\rgal$, $\mh/\mgal$, and $N\equiv M_{\rm gal}/m$. 
The first of these can be written
\beq
{r_t\over \rgal} = 2.0\times 10^{-9} 
\left({\rgal\over 1\ {\rm kpc}}\right)^{-1}
\left({\mh/m\over 10^6}\right)^{1/3}
\left({r_\star\over R_\odot}\right).
\eeq
In the ``empty loss cone'' limit, the tidal disruption rate 
varies only logarithmically with $r_t$ \citep{LS:77}.
The dependence on $r_t$ is steeper if the loss cone is partially
full (as is the case in these models); 
nevertheless, here we simply fix $r_\star=R_\odot$ and $m=M_\odot$.
Furthermore, half-light radii for dE galaxies average $\sim 1$ kpc
with a very weak dependence on galaxy luminosity (although
with considerable scatter; \cite{GG:03})
and so we set
\beq
{r_t\over \rgal} = 4.5\times 10^{-9} 
\left({N\over 10^{10}}\right)^{1/3}
\left({10^3 \mh\over M_{\rm gal}}\right)^{1/3}.
\label{eq:rt2}
\eeq
Equation~(\ref{eq:rt2}) gives the first of the three 
new parameters in terms of the other two.

We relate $N$ in this expression to observable galaxy 
properties by setting $m\approx M_\odot$ and expressing
$M_{\rm gal}$ in terms of galaxy absolute magnitude.
A typical B-band mass-to-light ratio for dE galaxies is
$2\lap (M/L)_B\lap 4$ in Solar units
(e.g. Seth et al. 2008) so the relation between 
$M_B$ and $N$ is roughly
\beq
\log_{10} N \approx 10 - 0.4\left(M_B+18\right).
\label{eq:NvsBT}
\eeq

In the evolutionary models presented above,
$N$ was a trivial parameter, affecting only how
the time unit of the calculation (essentially the
relaxation time) was converted into years
(e.g. eq.~\ref{eq:tr}).
In models containing a black hole,
$N$ also fixes the ratio between orbital periods 
$P(E)$ and the (orbit-averaged) time scales for diffusional 
loss-cone refilling, or
\begin{mathletters}
\begin{eqnarray}
q(E)&\equiv& \frac{1}{R_{lc}(E)} \oint 
\frac{dr}{v_r}
\lim_{R\rightarrow 0}
\frac{\langle(\Delta R)^2\rangle}{2R} \\
&\approx&  P(E)R_{lc}(E)^{-1} \tr^{-1}.
\label{eq:qofe}
\end{eqnarray}
\end{mathletters}
Here $R\equiv J^2/J_c(E)^2$ is a dimensionless
angular momentum variable, $0\le R\le 1$, with
$J_c(E)$ the angular momentum of a circular orbit
of energy $E$, and $\langle\left(\Delta R\right)^2\rangle$ is
the diffusion coefficient associated with $R$.
For a galaxy with given structural parameters ($\mgal,\rgal$),
$q\propto N^{-1}$.
Small/large $q$ (i.e. large/small $N$) correspond to empty/full
loss cone regimes respectively \citep{LS:77}.
We adopt the prescription of \cite{CK:78} for relating
the loss cone flux $\cal{F}$ to the phase-space density
$f$ given $q$.
The integrated loss rate, $\int {\cal F}(E) dE$, is denoted
by $\dot M$.
The contributions of the masses of tidally disrupted stars
either to the galactic potential or to the mass of the black
hole are ignored.

Masses of (putative) black holes in dE galaxies are poorly 
constrained, and so several values for the last of the three parameters
$\mh/M_{\rm gal}$ were tried, in the range 
$10^{-4}\le\mh/M_{\rm gal}\le 0.03$.

In order to accomodate a black hole, a
a slight modification to the equilibrium galaxy models
of equation~(\ref{eq:model}) was required.
In an isotropic stellar system containing a central point mass,
the central density must increase at least as fast as 
$r^{-1/2}$ in order for $f(E)$ to remain nonnegative.
The flat nuclear density profile of equation~(\ref{eq:model})
was therefore changed to a $\rho\propto r^{-0.5}$
dependence inside $r\approx r_c$.

In what follows, we fixed the following parameter values:
\beq
\gamma=4, \ \ \ n=3 .
\eeq
Setting $n=3$ is reasonable given the weak observed
dependence of galaxy Sersic index on $M_B$ for dE galaxies
\citep{GG:03}, but we note again that our models should not be scaled
to giant E galaxies which have steeper central density profiles.
We then carried out four series of integrations, 
starting from initial models with the following parameters:

\begin{itemize}

\item Series I: $M_{\rm nuc}/M_{\rm gal}=3\times 10^{-3}$,
$\mh/M_{\rm gal}=1\times 10^{-3}$.
The parameter $r_{\rm nuc}/r_{\rm gal}$ was varied
from $2\times 10^{-4}$ to $3\times 10^{-2}$.
The two mass ratios that define this series of models
are close to the average observed values
\citep{FF:05,ferrarese06a}.

\item Series II: $r_{\rm nuc}/r_{\rm gal}=3\times 10^{-4}$.
The nuclear mass was varied over the range
$1\times 10^{-4}\le M_{\rm nuc}/M_{\rm gal}\le 3\times 10^{-2}$
and the black hole mass was set to $M_{\rm nuc}/3$.

\item Series III: $M_{\rm nuc}/M_{\rm gal}=3\times 10^{-3}$,
$r_{\rm nuc}/r_{\rm gal}=3\times 10^{-4}$.
The black hole mass was varied over the range
$10^{-4} \le \mh/M_{\rm gal}\le 3\times 10^{-2}$.
 
\end{itemize}

\noindent
The number of stars in Series I-III was fixed at $N=10^{10}$,
appropriate for a $M_B\approx -18$ galaxy (eq.~\ref{eq:NvsBT}).
As discussed below, the choice of $N$ has very little effect
on the evolution of the density profile, and the results
for $\rho(r,t)$ can robustly be scaled to galaxies
of different luminosities after adjusting the value of $\tgal$.
The assumed value of $N$ does, however, substantially affect the 
dynamics of the loss cone and the rate of tidal disruptions, 
and to investigate this dependence 
a fourth series of integrations were carried out:

\begin{itemize}

\item Series IV: $M_{\rm nuc}/M_{\rm gal}=3\times 10^{-3}$,
$r_{\rm nuc}/r_{\rm gal}=1\times 10^{-3}$,
$\mh/M_{\rm gal}=1\times 10^{-3}$.
The parameter  $N$ was varied from $10^6$ to $10^{11}$;
according to equation~(\ref{eq:NvsBT}), this corresponds to a
range in galaxy absolute magnitude
$-8\gap M_B \gap -20.5$.
\end{itemize}

\noindent Each integration was continued until a time
of roughly $0.1 \tgal$.
Using equations~(\ref{eq:NvsBT}) and~(\ref{eq:leastsquares}),
this corresponds to 
$\sim 10^{14}$ yr for $M_B=-18$, 
$10^{13}$ yr for $M_B=-15$,
$10^{12}$ yr for $M_B=-12$, and
$10^{11}$ yr for $M_B=-9$.

\begin{figure}
\centering
\includegraphics[scale=0.65]{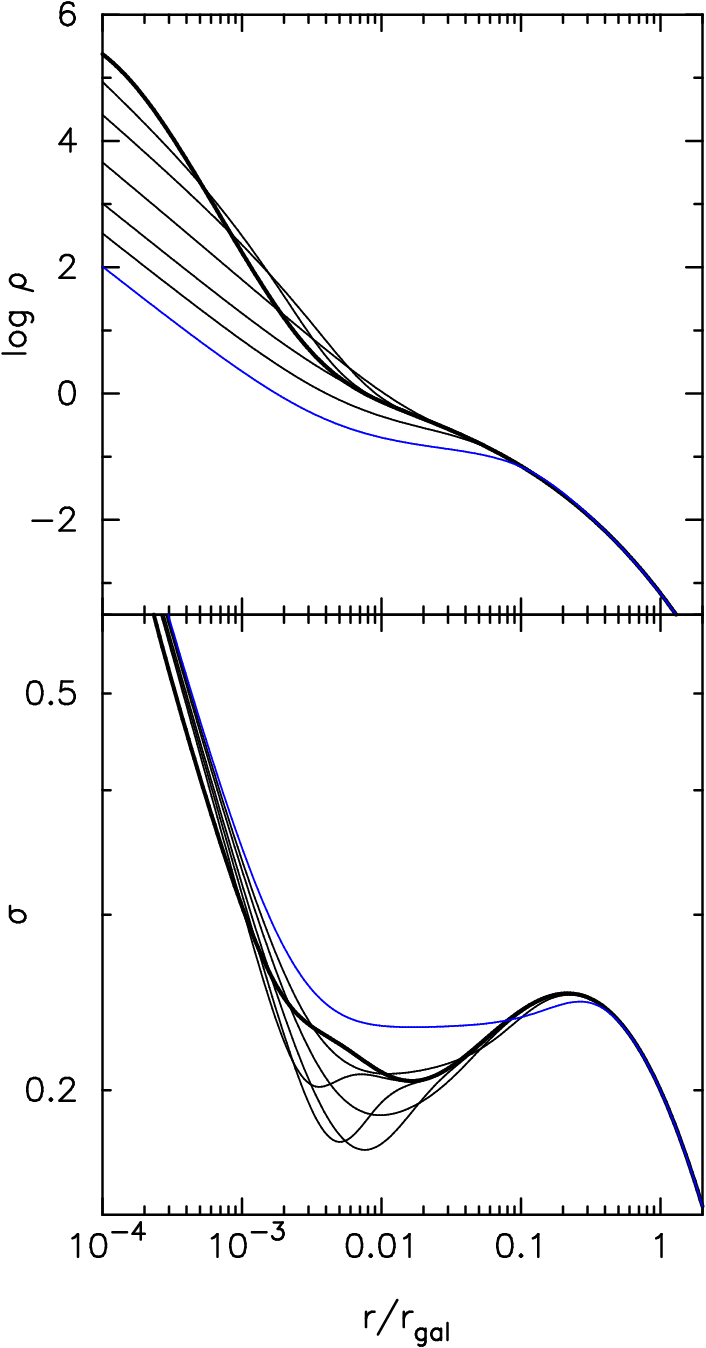}
\caption{Evolution of density (upper) and velocity dispersion (lower) 
profiles in a model from Series I with 
$M_{\rm nuc}/M_{\rm gal}=0.003$,
$r_{\rm nuc}/r_{\rm gal}=0.0002$ and
$\mh/M_{\rm gal} = 0.001$.
Thick curves are the initial models and blue curves show
the final time step;
in units of $\tgal$,
the displayed times are
$(0,3\times 10^{-6},1\times 10^{-5}, 1\times 10^{-4}, 1\times 10^{-3},
0.01, 0.1)$.
\label{fig:rhosigset3}}
\end{figure}

In a galaxy containing a massive black hole,
the nuclear half-mass relaxation time is not a well defined
quantity.
Instead, we will cite below the local relaxation time $t_r(r)$ in
these models,
defined as 
\beq
t_r(r) = {0.34 \sigma(r)^3\over\rho(r) m G^2\ln\Lambda}
\eeq
\citep{Spitzer:87}, where $\sigma(r)$ is the 1d
stellar velocity dispersion defined above and includes the
contribution to the stellar motions from the presence of
the black hole itself.
A natural radius at which to evaluate $t_r$ is the black
hole's influence radius,
\beq
\rbh = {G\mh\over\sigma_0^2}.
\eeq
In this formula, we set $\sigma_0$ equal to its peak value
in the black-hole-free model; e.g. for the models of
Figure~\ref{fig:two} ($n=3$), $\sigma_0 \approx 0.25$, 
or
\beq
\rbh = 0.016 \left({\mh\over 0.001}\right)
\eeq
in units where $G=M_{\rm gal}=1$.
This is reasonably equivalent to the way that $\rbh$ 
is defined by observers.
\cite{MMS:07} plot $\tr(\rbh)$ as a function of $\sigma_0$
in a sample of (mostly luminous) galaxies; they find 
$\tr(\rbh)\approx 10^{11}\ {\rm yr} (\sigma_0/100\ {\rm km\ s}^{-1})^{7.5}$.

\begin{figure}
\centering
\includegraphics[scale=0.70]{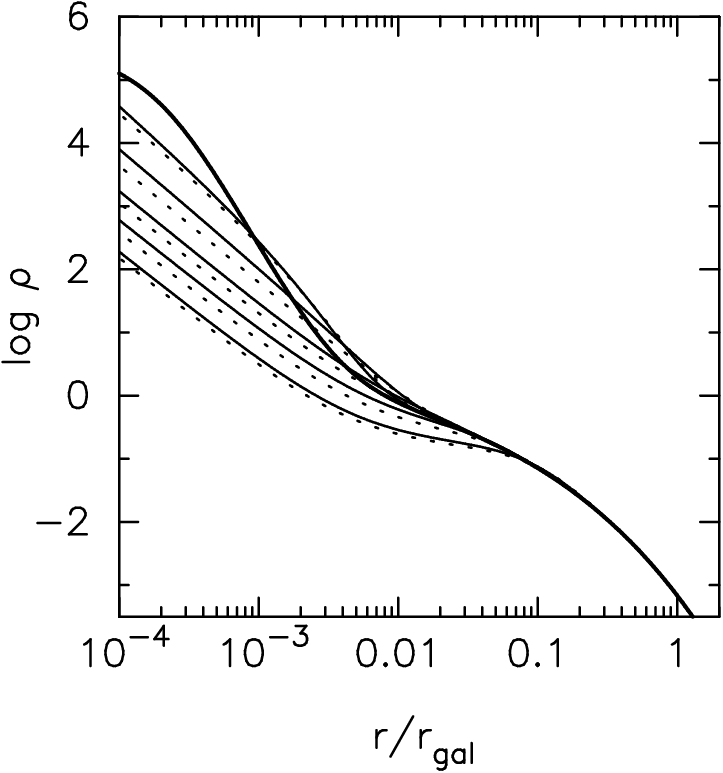}
\caption{Comparison of the density evolution in two
integrations including (solid lines) and without (dashed lines)
the loss term ${\cal F}$ representing tidally disrupted
stars.
Initial model (heavy line) had 
$M_{\rm nuc}/M_{\rm gal}=0.003$,
$r_{\rm nuc}/r_{\rm gal}=0.0003$ and
$\mh/M_{\rm gal} = 0.001$.
In units of $\tgal$,
the displayed times are
$(0,1\times 10^{-5},1\times 10^{-4}, 1\times 10^{-3},0.01, 0.1)$.
\label{fig:compareset3}}
\end{figure}

As defined here, $\rbh$ is essentially time-independent in
these simulations, although the values of $\rho(\rbh)$,
$\sigma(\rbh)$ and hence $\tr(\rbh)$ do evolve substantially.

\subsection {Results: Density Evolution}

Models with central black holes evolve differently than
models without black holes, in two important ways.

\medskip
\noindent
1. On time scales of order the nuclear relaxtion time,
the density within the black hole's influence sphere 
attains the $\rho\sim r^{-7/4}$ ``zero-flux'' Bahcall-Wolf 
quasi-steady-state form \citep{BW:76,BW:77}.

\medskip
\noindent
2. The presence of the black hole counteracts
both the tendency toward core collapse in compact nuclei, 
and the tendency toward core expansion in low-density nuclei.
The black hole's mass -- while a small fraction of the
total galaxy mass -- is always comparable with the mass of
the nuclear cluster in these models, 
and the fixed gravitational potential associated
with the black hole acts to decouple changes in the central density
from changes in the central velocity dispersion, 
breaking the feedback loop that drives core collapse.
Furthermore, in low-density nuclei, the presence of the black 
hole raises the ``temperature'' of the nucleus and decreases the 
inward heat flux from the surrounding galaxy, 
reducing the expansion rate compared
with that of a black-hole-free nucleus.

\medskip
\noindent
Because core collapse is suppressed in these models, the
long-term evolution ($t\gap t_r(\rbh)$) is always toward expansion.

\begin{figure}
\centering
\includegraphics[scale=0.50]{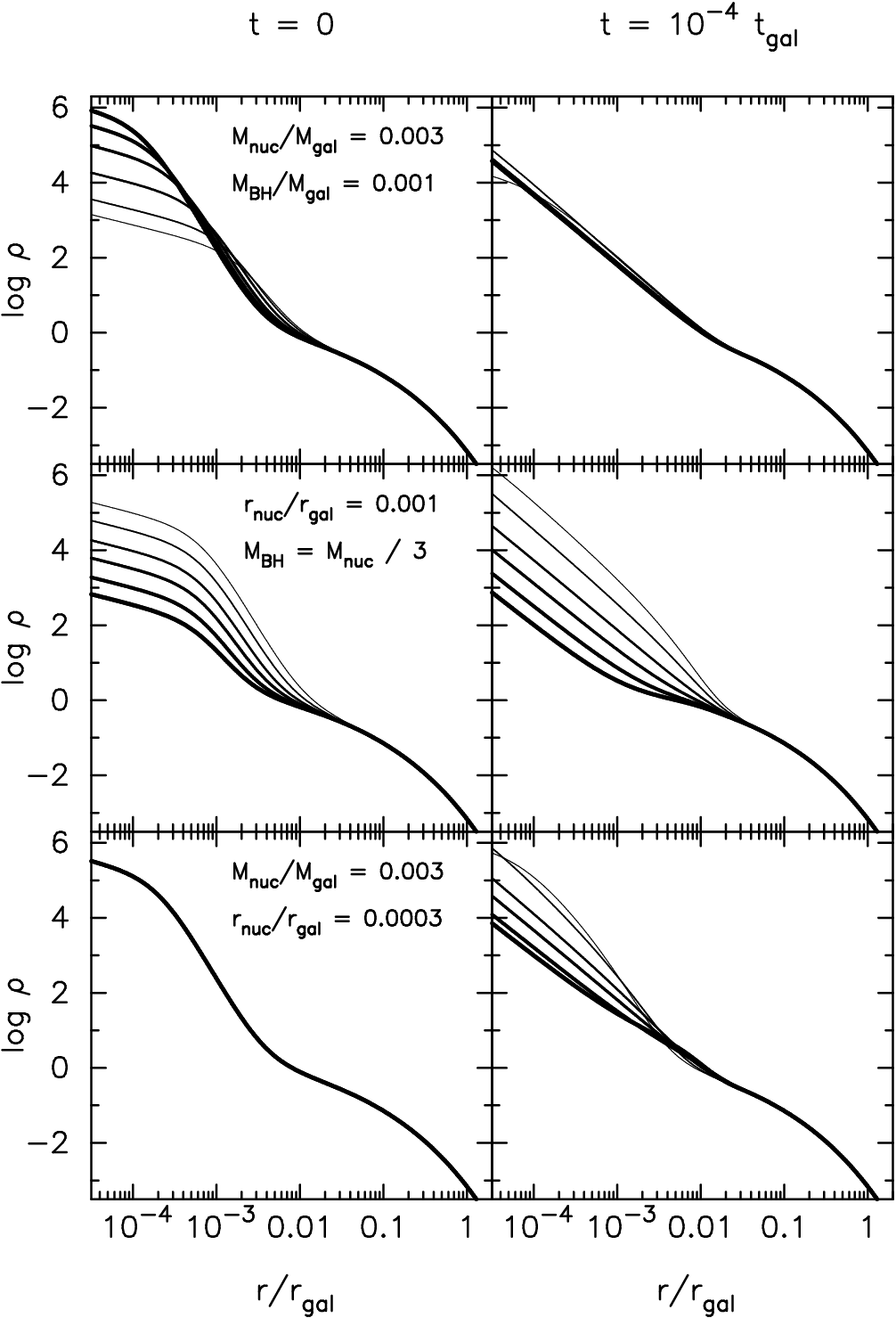}
\caption{Density profiles at $t=0$ (left) and after
an elapsed time of $10^{-4}\tgal$
(right) for various models containing black holes.
Line widths correspond between left and right panels.
Top: Series I; $r_{\rm nuc}/r_{\rm gal}=
(2\times 10^{-4},3\times 10^{-4},1\times 10^{-3},3\times 10^{-3},
1\times 10^{-2},3\times 10^{-2})$;
line width is a decreasing function of $r_{\rm nuc}$.
Middle: Series II; $M_{\rm nuc}/M_{\rm gal}=
(1\times 10^{-4},3\times 10^{-4},1\times 10^{-3},3\times 10^{-3},
1\times 10^{-2}, 3\times 10^{-2})$;
line width is a decreasing function of $M_{\rm nuc}$.
Bottom: Series III; 
$\mh/M_{\rm gal} = 1\times 10^{-4}, 3\times 10^{-4}, 1\times 10^{-3},
3\times 10^{-3}, 1\times 10^{-2}, 3\times 10^{-2}$;
line width is a decreasing function of $\mh$.
\label{fig:initialfinal}}
\end{figure}

Figure~\ref{fig:rhosigset3} illustrates the evolution of a model
from Series I, with 
$M_{\rm nuc}/M_{\rm gal}=0.003$,
$r_{\rm nuc}/r_{\rm gal}=0.0002$ and
$\mh/M_{\rm gal} = 0.001$.
On a time scale of $\sim\tr(\rbh)\approx 10^{-5}\tgal$
the Bahcall-Wolf cusp 
is established near the black hole and the velocity
dispersion drops slightly due to the change in the density
slope.
At longer times the inward flux of
heat from the galaxy begins to erase the dip in the velocity
dispersion profile and causes the density to drop.

In addition to heat transfer from the galaxy, 
destruction of stars by the black hole is 
effectively a heat source which by itself would cause
the density of the nucleus to fall \citep{Shapiro:77}.
However, integrations in which the loss term ${\cal F}(E,t)$ was
set to zero evolved almost
identically to those which included the loss term.
Figure~\ref{fig:compareset3} shows an example. 
Apparently, in all these models, the difference
in temperature between nucleus and galaxy, and not
the destruction of stars by the black hole,
is the dominant effect which causes the nucleus to expand.

\begin{figure}
\centering
\includegraphics[scale=0.50]{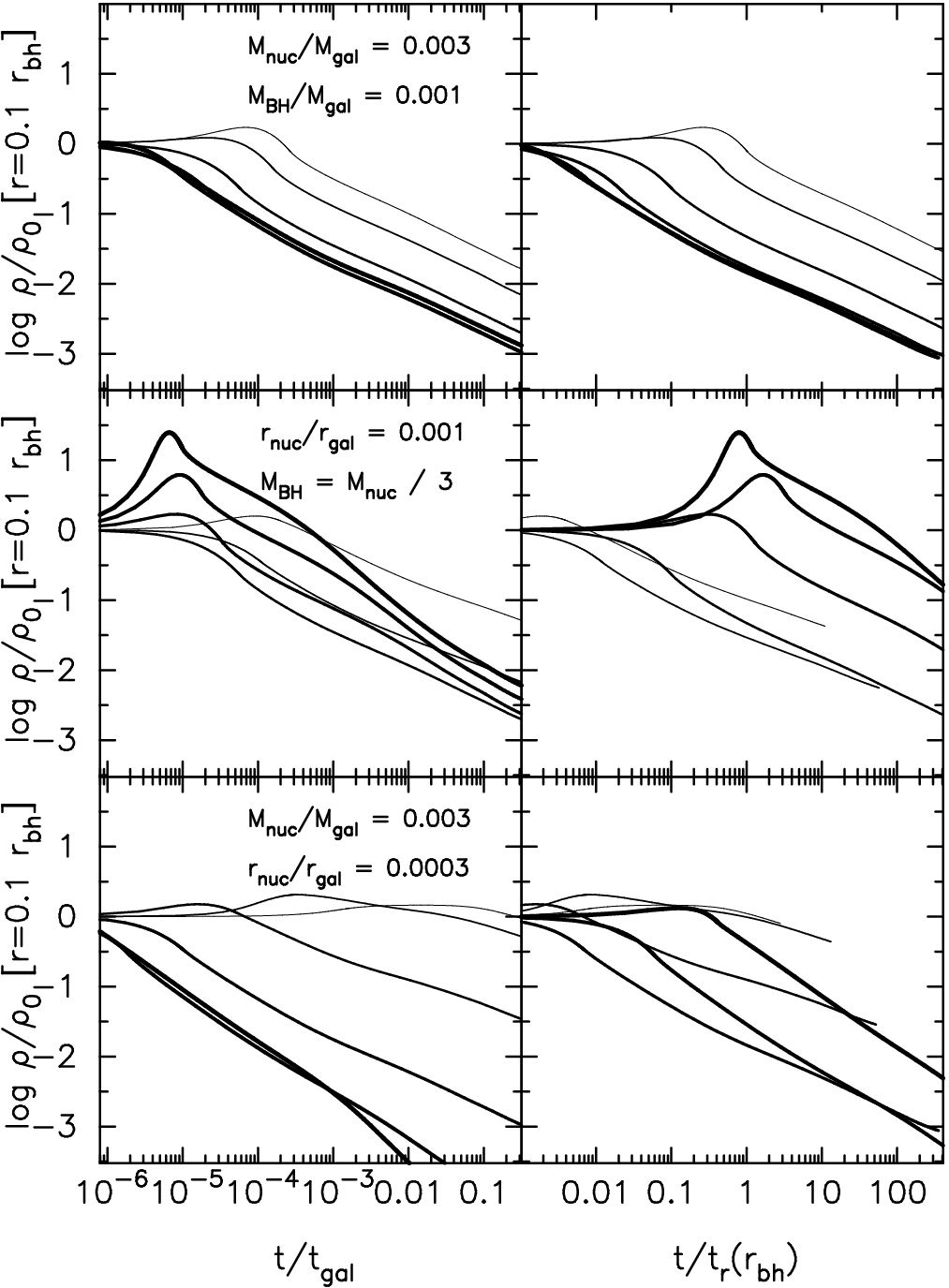}
\caption{
Evolution of the central density in the black hole
models illustrated in Fig.~\ref{fig:initialfinal}.
The vertical axis is the density at a radius of
$0.1$ times the black hole's initial influence radius,
compared with its value at $t=0$.
Left: time in units of galaxy half-mass relaxation time.
Right: time in units of the galaxy's initial relaxation
time evaluated at the black hole's influence radius.
\label{fig:evolbh}}
\end{figure}

Figure~\ref{fig:initialfinal} illustrates the early
evolution of models from Series I-III;
the figure shows density profiles at $t=0$, and 
after an elapsed time of $10^{-4} t_{\rm gal}$.
The top row of the figure shows the effects of varying the nuclear
radius in galaxies with $M_{\rm nuc}/M_{\rm gal}=0.003$
and fixed black hole mass (Series I).
In each of these models, the initial velocity dispersion profile
exhibits a dip outside $r_{bh}\approx 0.015$, 
and the early evolution
is a combined result of the inward flow of heat from the galaxy,
and formation of the Bahcall-Wolf profile at $r\lap r_{bh}$
due to scattering of stars onto tightly-bound orbits.
By $\sim 3\times 10^{-5} \tgal$ almost all differences between
the initial profiles have been erased and the nucleus
has been transformed into a $\rho\sim r^{-7/4}$ cusp
with a mass approximately equal to the initial nuclear mass.

The middle row of Figure~\ref{fig:initialfinal}
shows the result of varying the nuclear mass at fixed
nuclear radius; the black hole mass is set to $M_{\rm nuc}/3$
(Series II).
In low-mass nuclei, the evolution is similar to that of
the models in Series I, with the difference that the final
normalization of the Bahcall-Wolf cusp is proportional
to the initial nuclear mass.
For $M_{\rm nuc}\gap 0.01 M_{\rm gal}$ ($\mh\gap 0.003 M_{\rm gal})$,
the initial velocity dispersion profile lacks a local minimum
and the heat flow from the galaxy is greatly reduced; 
the evolution in these models is dominated by formation 
of the Bahcall-Wolf cusp.

The bottom row of Figure~\ref{fig:initialfinal}
shows the effects of varying the black hole mass
on a model with fixed $(M_{\rm nuc}, r_{\rm nuc}$) (Series III).
Here the differences are due almost entirely to the effect of
the black hole on the velocity dispersion profile:
larger black holes imply a smaller, initial dip in the 
temperature profile and less evolution.
In these models, the contribution of the loss term
to the changes in $\rho(r)$ is most clearly seen, but it is
still slight.
It is interesting to note that the form of $\rho(r)$ 
at $r<r_{bh}$ can depart strongly from the ``zero-flux,''
$\rho\sim r^{-7/4}$ form at early times in the models
with larger black holes.

Evolution of the central density over the full course of the
integrations is shown in Figure~\ref{fig:evolbh} for the same models in
Figure~\ref{fig:initialfinal}.
Plotted there is the density at a (fixed) radius of $0.1 r_{bh}$,
normalized to its initial value.
Aside from an early increase due to the formation of the
$\sim r^{-7/4}$ cusp, the general trend is for
the density to decrease;
the decrease is smallest (all else equal) for bigger
black holes, for the reasons discussed above.
For $\mh/M_{\rm gal}=0.001$, the density declines
as $\rho\sim t^{-0.5}$ at late times ($t\gap t_r(r_{bh})$),
with steeper(shallower) falloff for smaller(larger) $\mh$.

What do these models imply for the observed properties
of nuclear star clusters?
The general tendency of the black hole models is
to evolve toward lower central density at fixed
nuclear size, i.e. to move toward the 
left on Figure~\ref{fig:nuclei}, and eventually to
disappear from this plot as the nuclei become too
diffuse to be detected.
Evolution is strongest in the case of models containing
nuclei of low initial density and small 
($\mh \lap 3\times 10^{-4}M_{\rm gal}$)
black holes: these galaxies have a strong ``temperature
inversion,'' implying nuclear heating,
while the presence of the black hole inhibits core collapse.
The heavy curves in the lower left panels of 
Figures~\ref{fig:initialfinal} and~\ref{fig:evolbh}
show that the nuclei in such models are the most rapidly
destroyed, and assuming the presence of such ``small'' black 
holes in nucleated galaxies
would help to explain the lack of points in the
upper left part of Figure~\ref{fig:nuclei}.
Larger black holes ($\mh\gap M_{\rm nuc}$) tend to 
stabilize nuclei by reducing the temperature inversion and
decreasing the flow of heat from the galaxy.
In such models, a Bahcall-Wolf cusp is established 
in a time $\sim\tr(\rbh)\approx 10^{-5}\tgal$
and little evolution subsequently takes place.

While no attempt was made to construct evolutionary black-hole
models that match the current state of NGC 205, 
it is clear from Figures~\ref{fig:initialfinal} 
and~\ref{fig:evolbh} that including a black hole of mass 
$\mh\gap 10^{-3}\mgal$ in this galaxy would cause the
initial configuration to evolve very little over $10$ Gyr,
aside from the establishment of the $\rho\sim r^{-7/4}$ cusp
inside $\sim 0.2\rbh\approx 0.1\ {\rm pc} (\mh/10^5\msun)$.
The kinematical upper limit to $\mh$ in NGC 205 is
$\sim 5\times 10^4\msun$ \citep{Valluri:05} so the influence of 
a black hole would be limited to $\sim 0.05$ pc $\approx 0.02''$, 
below HST resolution.

\begin{figure*}
\centering
\includegraphics[scale=0.9]{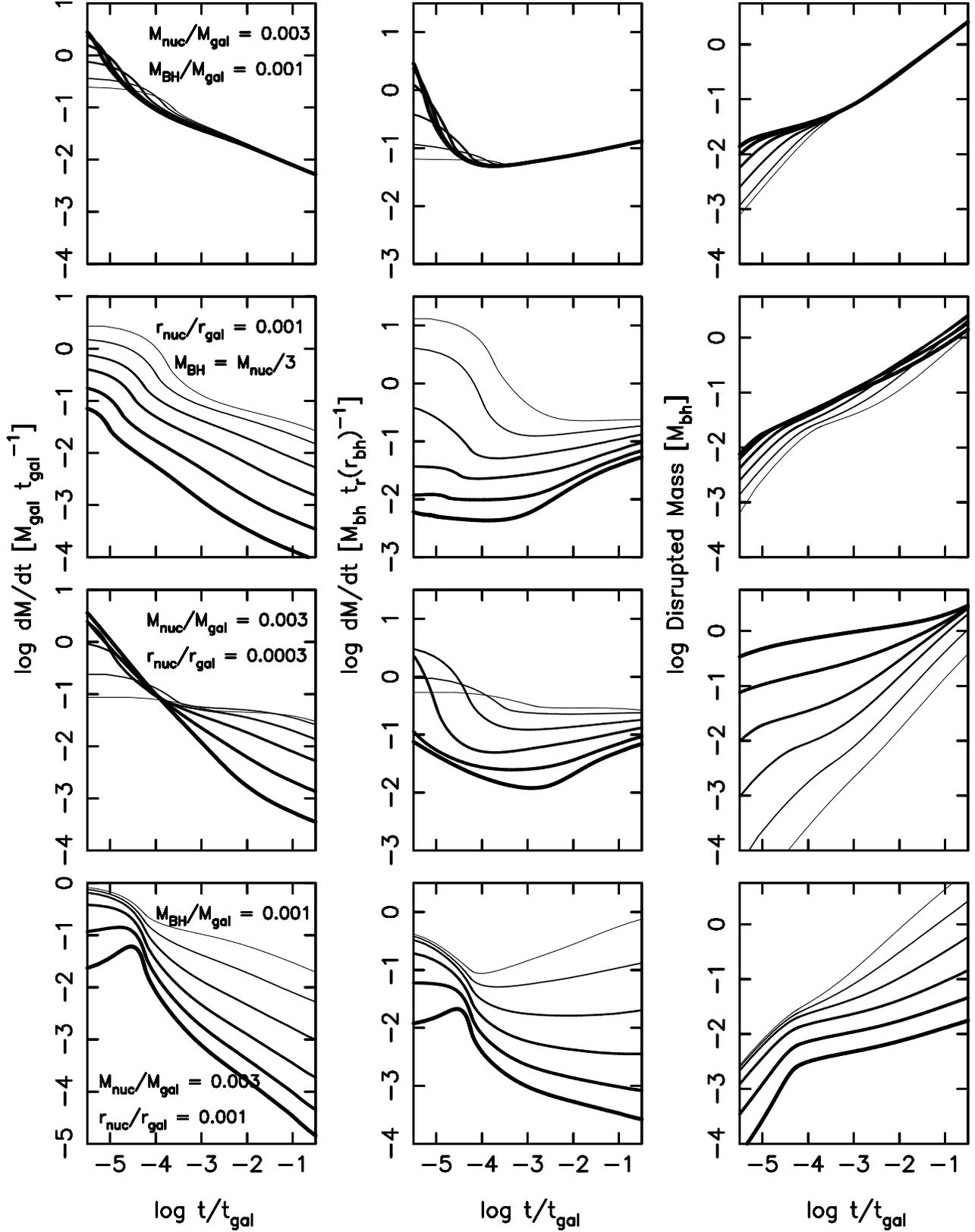}
\caption{{\it Left column:} Rate at which mass is
scattered into the tidal disruption sphere of the black hole,
expressed in units of the galaxy mass divided by the galaxy
half-mass relaxation time (both of which are constant).
{\it Middle column:} Same as the left column, except that
the loss rate is expressed in units of black hole mass divided
by the (time-varying) relaxation time at the
black hole influence radius $r_{bh}$.
{\it Right column:} The integral until time $t$ of $dM/dt$,
i.e. the total mass scattered into the black hole's
disruption sphere, in units of the (constant) black hole mass.
Lines and line widths correspond to the same models
in Figures~\ref{fig:initialfinal} and~\ref{fig:evolbh}.
The bottom panel shows, in addition, models from Series IV, with 
$\log_{10}N=(6,7,8,9,10,11)$; line width decreases with
increasing $N$.
\label{fig:rates}}
\end{figure*}

\begin{figure}
\centering
\includegraphics[scale=0.55]{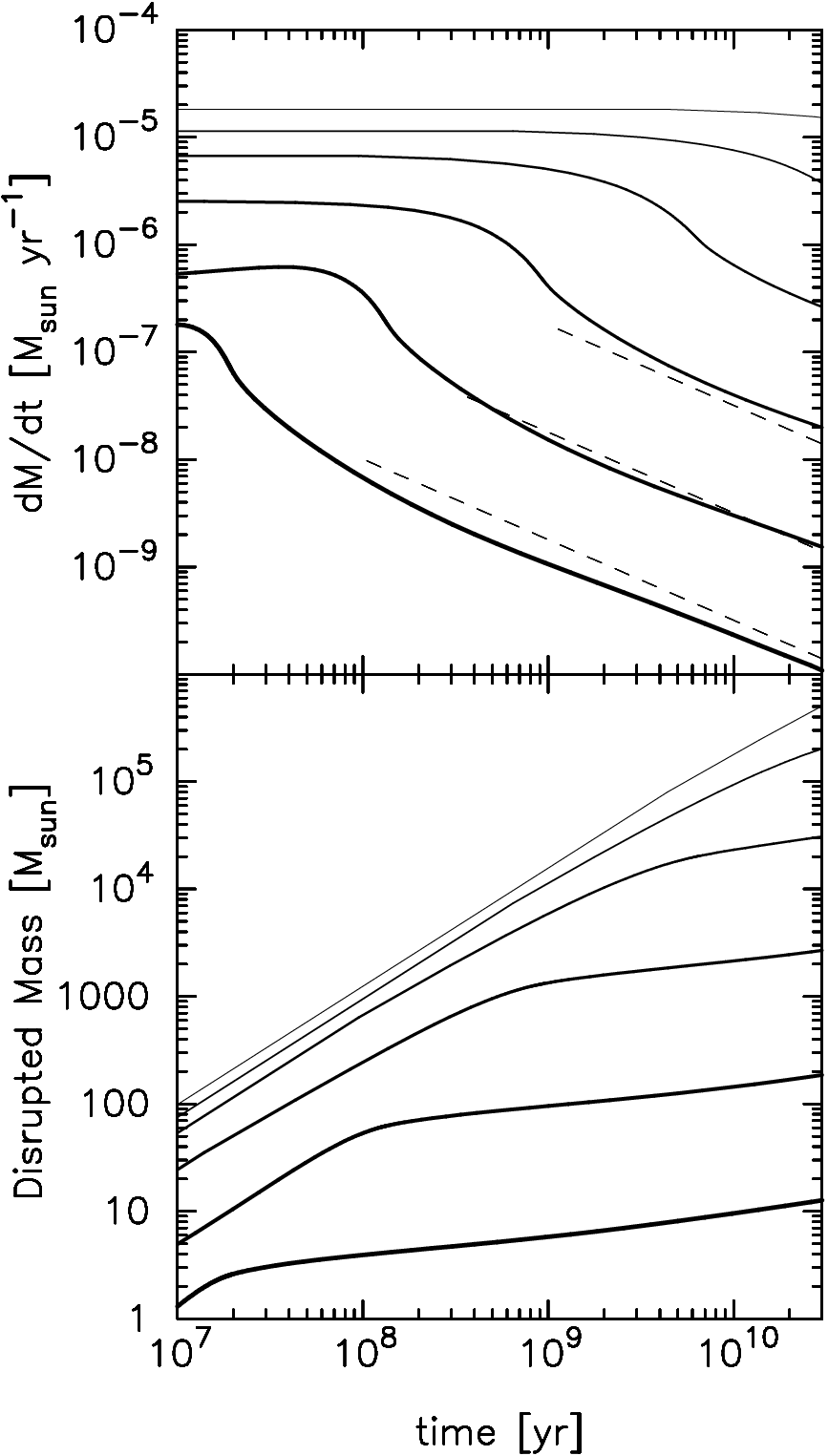}
\caption{{\it Top:} Stellar tidal disruption rate
vs. time for models from Series IV,
which have fixed initial structural parameters but
which vary in their assumed total mass $\mgal$ and 
hence in the number of stars, $N=\mgal/\msun$.
Curves correspond to $N$ values ranging from $10^6$ 
(thickest) to $10^{11}$ (thinnest) increasing in steps 
of 10;
the corresponding galaxy luminosities range from
$M_B=-8$ to $M_B=-20.5$.
Black hole masses are fixed at $10^{-3}\mgal$.
Dashed lines are defined in the text (eq.~\ref{eq:dashed}).
{\it Bottom:} Total disrupted mass.
Galaxy relaxtion times have been assumed to scale with
galaxy  luminosity as in eq.~\ref{eq:leastsquares}.
\label{fig:rateevol}}
\end{figure}

\begin{figure*}
\centering
\includegraphics[scale=0.90]{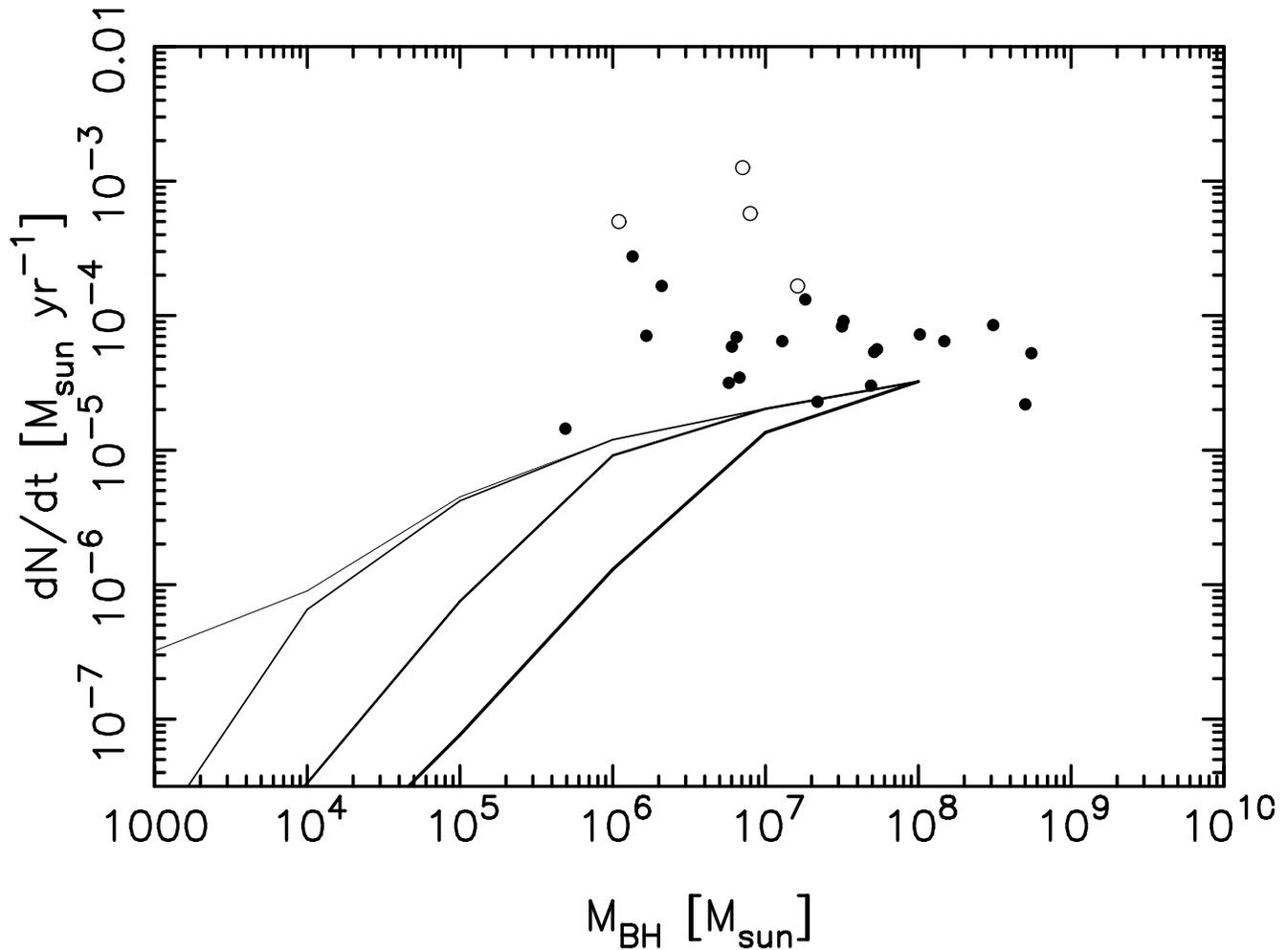}
\caption{Points are stellar tidal disruption rates
computed by Wang \& Merritt (2004) for a sample of
early-type, ``power-law'' galaxies, with luminosity
profiles parametrized as ``Nuker'' laws.
Open circles are galaxies for which the inner (projected)
slope parameter $\Gamma$ of the Nuker-law fit was 
0.9 or greater, corresponding to a $\rho\sim r^{-2}$
spatial density profile for the inner parts of the galaxy.
Lines show disruption rates in the Fokker-Planck
models from Series IV (Fig.~\ref{fig:rateevol})
at four times, $0.01$ Gyr (thin line), $0.1$ Gyr,
$1$ Gyr and $10$ Gyr (thick line).
\label{fig:wang}}
\end{figure*}

\subsection{Results: Rates of Stellar Tidal Disruption}

An unambiguous signature of a massive black hole is a 
tidal disruption flare \citep{KB:99}.
Figure~\ref{fig:rates} shows the rate
at which mass (in stars) is scattered into the black
hole's disruption sphere for the various models.
The left and middle columns of this figure show $\dot M$ 
expressed in units of $M_{\rm gal}/t_{\rm gal}$ 
(which does not change with time) and 
$\mh/t_r(r_{bh})$ respectively;
the latter unit is itself a function of time
via the time dependence of $\rho(\rbh,t)$ and $\sigma(\rbh,t)$.
(We emphasize that the mass of the black hole is assumed
constant in these Fokker-Planck models.)
The right column in Figure~\ref{fig:rates} shows the total
mass lost to disruptions, integrated from time zero until
time $t$, in units of $\mh$.

While the variation of mass loss rate with time is complex,
the general trend is for $\dot M$ to decrease
as the nuclear density drops.
We consider first a ``fiducial'' model defined by 
$\mnuc/\mgal\approx0.003$ and $\mh/\mgal\approx 0.001$;
these are approximately equal to the average mass ratios
that are observed in galaxies with nuclear clusters and
black holes respectively \citep{wehner06,ferrarese06a}.
We also focus first on the case $N=10^{10}$ which was assumed
in Series I-III;
this corresponds to a galaxy absolute magnitude
$M_B\approx -18$ (eq.~\ref{eq:NvsBT}) and to 
$\tgal\approx 7\times 10^{14}$ yr (eq.~\ref{eq:leastsquares}).
By $t= 2\times 10^{-5}\tgal\approx 15$ Gyr, 
the Bahcall-Wolf cusp has been established in 
these models and the loss rate to the
black hole is essentially independent of the initial nuclear
radius:
\beq
{\dot M}\approx 0.1 {\mh\over t_r(r_{bh})}.
\eeq
$\dot M$ remains relatively constant thereafter if expressed in
these units.
The total mass in disrupted stars at $t\approx 2\times 10^{-5}\tgal$
in the same subset of models is $\sim 0.02\mh$, implying a mean
disruption rate over this time of $\sim 10^{3}\mh/\tgal$,
or
\beq
\overline{\dot M}\approx 10^{-6} \msun\ {\rm yr}^{-1}
\left({\mh\over 10^6\msun}\right)
\left({\tgal\over 10^{14} {\rm yr}}\right)^{-1}.
\eeq

However, much higher and much lower rates are also possible:

\noindent
-- At early times, before the Bahcall-Wolf cusp is
fully established,
$\dot M$ in the fiducial models is higher for smaller $\rnuc/\rgal$
by as much as $\sim$ an order of magnitude.

\noindent
-- The second row of Figure~\ref{fig:rates} shows that the 
initial $\dot M$ scales roughly with $\mnuc$ for fixed
$\rnuc/\rgal$ and $\mh/\mnuc$.

\noindent
-- The third row of Figure~\ref{fig:rates} shows that the
mass loss rate at early times (i.e. before the nuclear
density profile has been strongly affected by the presence
of the black hole) is a decreasing function of $\mh$
at fixed ($\mgal,\tgal$),
for reasons that were discussed in detail by \cite{WM:04}.

The results discussed so far have described the effects
on $\dot M$ of changing the structural parameters
of the initial galaxy.
An equally important parameter is the assumed number
of stars $N$.
In addition to fixing the unit of time $\tgal$,
$N$ also determines the details of the loss-cone refilling
process and hence the number of stars scattered into the
black hole in one relaxation time.
Small $N$ corresponds to a ``full loss cone''
and large $N$ to an ``empty loss cone.''
(Small values of $\mh$ also imply a full loss cone;
for instance, in the third row of Fig.~\ref{fig:rates},
the full-loss-cone regime is reached for $\mh/\mgal\approx 10^{-3}$.)

Recalling the definition of $q(E)$ (eq.~\ref{eq:qofe})
as the dimensionless ratio between the orbital period and the 
loss cone refilling time, these two extremes correspond to
$q\gg 1$ and $q\ll 1$ respectively.
It can be shown that the number of stars scattered into 
the loss cone in one relaxation time scales with $q$ as
\begin{mathletters}
\begin{eqnarray}
{\cal F}\times \tr
&\propto& \left[\ln J_c^2/J_{lc}^{2}\right]^{-1},\ \ q\ll 1; \\
&\propto& q^{-1} \propto N,\ \ \ \ \ \ \ \ \ \ q\gg 1
\end{eqnarray}
\end{mathletters}
where the $q\propto N^{-1}$ dependence assumes a galaxy with fixed
structural properties, i.e. fixed ($\mgal,\rgal$).
This decrease in $\dot M\tgal$ at small $N$ is visible
in the last row of Figure~\ref{fig:rates};
for $N\lap 10^8$, i.e. $M_B\gap -13$, the loss cones in these
models are essentially full and $\dot M$ is independent of
the relaxation time.

The $N$-dependence is illustrated more clearly in Figure~\ref{fig:rateevol}
which shows $\dot M$ and $\int {\dot M}dt$ for the same models,
expressed this time in units of solar masses and years;
equations~(\ref{eq:leastsquares}) and (\ref{eq:NvsBT})
were used to convert $N$ and $\tgal$ into physical units.
Each of these models evolves in nearly the same way
when expressed in terms of $t/\tgal$; as discussed above,
changing the flux of stars into the loss cone has very
little effect on the density evolution.
The differences in the curves of Figure~\ref{fig:rateevol}
are due primarily to the different time scaling (small
galaxies evolve faster) and secondarily to changes in
$q(E)$ as $\rho(r)$ evolves.

The initial disruption rate scales almost linearly 
with galaxy mass.
This dependence is most easily understood in the 
small-$N$/full-loss-cone limit, where 
$\dot M\propto \mgal/\tcr\propto 
(\mgal/\trh)(\trh/\tcr)$ with $\tcr$ the galaxy crossing time.
From Figure~\ref{fig:nuclei}a, $\trh\sim \mgal$,
and $\trh/\tcr\sim N\sim\mgal$, so $\dot M\sim\mgal\sim N$.
This is similar to the dependence reported in other studies,
e.g. Figure~21 of \cite{freitag06} who considered black holes
in power-law nuclei.

After the Bahcall-Wolf cusp is established,
the evolution of $\dot M$ is dominated by the systematic
decrease in nuclear density.
The dashed lines in the top panel of Figure~\ref{fig:rateevol},
which accurately reproduce the disruption rates for
$t\gap \tr(\rbh)$, are
\beq
\dot M = 1.8\times 10^{-7} \msun {\rm yr}^{-1} 
\left({\mgal\over 10^8\msun}\right) 
\left({t\over 10^9 {\rm yr}}\right)^{-3/4}.
\label{eq:dashed}
\eeq
The bottom  panel shows time-integrated masses;
these are roughly $10^{-2}\mh$ at $t=10$ Gyr in all the models.

Tidal destruction rates computed in the past 
for luminous E galaxies have generally been found to 
decrease with increasing galaxy/black hole mass
\citep{SU:99,MT:99,WM:04}, opposite to the trend
shown here in Figure~\ref{fig:rateevol}.
One reason is probably the rather different models used
in these studies for representing $\rho(r)$, 
e.g. ``Nuker'' models which have unbroken power-law
cusps at small radii \citep{Nuker}.
These are reasonable models for some
galaxies but do not reproduce the nuclear
star clusters of low-luminosity ellipticals
 \citep{ferrarese06b}.
Figure~\ref{fig:wang} plots
disruption rates from \cite{WM:04} for a sample of 
galaxies modelled with Nuker profiles,
compared with $dN/dt$ from the Series IV models
at four different times: 0.01, 0.1, 1 and 10 Gyr.
The rates computed here at $t=0$ trace the lower envelope
of the points from \cite{WM:04}.

While these results suggest that tidal disruption
rates for some of the galaxies modelled by \cite{WM:04}
might have been over-estimated, such a conclusion is
probably premature until detailed models can be 
constructed based on the newest luminosity profiles
for these galaxies.
Six of the galaxies in the Wang \& Merritt sample
are classified as ``nucleated'' (Type Ia or Type Ib)
by \cite{ACS8}.
Inspection of the ACS luminosity profiles of these
galaxies suggests that none of them exhibit a clear,
two-component structure like that of NGC 205;
a ``Nuker'' model might therefore provide an adequate
fit to these galaxies.

Figure~\ref{fig:wang} also emphasizes the strong time dependence
of $\dot N$ in low-luminosity galaxies, due to the structural
changes caused primarily by transfer of heat from the galaxy.
This plot suggests that tidal flaring rates in dE galaxies are 
likely to be a strong function of the time since nuclear formation.

\section{Summary}

1. Half-mass relaxation times in nuclear star clusters
are interestingly short, ranging from $\sim 10$ Gyr
in galaxies with absolute magnitude $M_B=-18$ to 
$\sim 1$ Gyr at $M_B=-15$.

2. In the absence of a massive black hole, evolution
of nuclear star clusters is a competition between core
collapse and heating from the external galaxy.
Observed nuclei appear to be in the core-collapse
regime, with the exception of nuclei in the brightest
galaxies whose relaxation times are so long that they 
would not evolve appreciably in $10$ Gyr.

3. Adding a massive black hole ``turns off'' core collapse,
and nuclear star clusters containing black holes always
expand, due primarily to heat input from the galaxy, 
but also due to the effective heating associated with
stellar tidal disruptions.
Nuclei with modest black holes dissipate more
quickly than black-hole-free nuclei;
large black holes, on the other hand, tend to slow the transfer of heat
from the galaxy by reducing the temperature gradient.

4. Rates of stellar tidal disruption generally decrease
with time due to the drop in density of the nucleus.
Rates are lower in galaxies of lower mass,
assuming that the relation between black hole mass
and bulge mass extends to low galaxy luminosities.
A simple expression is presented that reproduces the time
dependence of the disruption rate in a restricted
class of models.

\acknowledgements

I thank Dan Batcheldor for assistance with 
the HST point-spread function used in the analyis of the NGC 205
luminosity profile,
Laura Ferrarese and Monica Valluri for additional advice on the
modelling of this galaxy,
and Milos Milosavljevic and Hagai Perets for useful discussions
about the dynamics.
Stefanie Komossa also provided useful comments.
This work was supported by grants 
AST-0420920 and AST-0437519 from the 
NSF, grant 
NNG04GJ48G from NASA,
and grant HST-AR-09519.01-A from
STScI.

\end{document}